\documentclass{article}

\usepackage{microtype}
\usepackage{graphicx}
\usepackage{subfigure}
\usepackage{booktabs} % for professional tables
\usepackage{amsmath}
\usepackage{amssymb}
\usepackage{balance}
\usepackage{hyperref}

\usepackage{relsize}

\usepackage[accepted]{icml2019}

\icmltitlerunning{FloWaveNet : A Generative Flow for Raw Audio}

\begin{document}

\twocolumn[
\icmltitle{FloWaveNet : A Generative Flow for Raw Audio}



\icmlsetsymbol{equal}{*}

\begin{icmlauthorlist}
\icmlauthor{Sungwon Kim}{snu}
\icmlauthor{Sang-gil Lee}{snu}
\icmlauthor{Jongyoon Song}{snu}
\icmlauthor{Jaehyeon Kim}{kakao}
\icmlauthor{Sungroh Yoon}{snu,asri}
\end{icmlauthorlist}

\icmlaffiliation{snu}{Electrical and Computer Engineering, Seoul National University, Seoul, Korea}
\icmlaffiliation{kakao}{Kakao Corporation}
\icmlaffiliation{asri}{ASRI, INMC, Institute of Engineering Research, Seoul National University, Seoul, Korea}
\icmlcorrespondingauthor{Sungroh Yoon}{sryoon@snu.ac.kr}

\icmlkeywords{Machine Learning, ICML}

\vskip 0.3in
]



\printAffiliationsAndNotice{}
\begin{abstract}
Most modern text-to-speech architectures use a WaveNet vocoder for synthesizing high-fidelity waveform audio, but there have been limitations, such as high inference time, in its practical application due to its ancestral sampling scheme. The recently suggested Parallel WaveNet and ClariNet have achieved real-time audio synthesis capability by incorporating inverse autoregressive flow for parallel sampling. However, these approaches require a two-stage training pipeline with a well-trained teacher network and can only produce natural sound by using probability distillation along with auxiliary loss terms. We propose FloWaveNet, a flow-based generative model for raw audio synthesis. FloWaveNet requires only a single-stage training procedure and a single maximum likelihood loss, without any additional auxiliary terms, and it is inherently parallel due to the characteristics of generative flow. The model can efficiently sample raw audio in real-time, with clarity comparable to previous two-stage parallel models. The code and samples for all models, including our FloWaveNet, are publicly available.

\end{abstract}

\section{Introduction} \label{intro}

The end-to-end waveform audio synthesis model is a core component of text-to-speech systems. The striking success of deep learning based raw audio synthesis architectures has led to modern text-to-speech systems leveraging them as vocoders to synthesize realistic waveform signals that are nearly indistinguishable from natural sound in the real world.

Current state-of-the-art text-to-speech architectures commonly use the WaveNet vocoder with a mel-scale spectrogram as an input for high-fidelity audio synthesis \cite{shen2018natural, arik2017deep1, arik2017deep2, ping2017deep, jia2018transfer}. However, the practical application of WaveNet has been limited because it requires an autoregressive sampling scheme, which serves as a major bottleneck in real-time waveform generation.

Several variations of the original WaveNet have been proposed to overcome its slow ancestral sampling. Parallel WaveNet \cite{oord2017parallel} has achieved real-time audio synthesis by incorporating inverse autoregressive flow (IAF) \cite{kingma2016improved} for parallel audio synthesis. The recently suggested ClariNet \cite{ping2018clarinet} presented an alternative formulation by using a single Gaussian distribution with a closed-form Kullback-Leibler (KL) divergence, contrasting with the high-variance Monte Carlo approximation from Parallel WaveNet.

Despite the success of real-time high-fidelity waveform audio synthesis, all of the aforementioned approaches require a two-stage training pipeline with a well-performing pre-trained teacher network for a probability density distillation training strategy. Furthermore, in practical terms, these models can synthesize realistic audio samples only by using additional auxiliary losses. For example, if only probability density distillation loss is used, Parallel WaveNet is prone to mode collapse, in which the student network converges to a certain mode of the teacher distribution, resulting in sub-optimal performance \cite{oord2017parallel}.
%% 교정에서 the realistic audio sample --> a realistic audio sample 로 수정이었는데 걍 realistic audio samples로 바꿈 // 위에서 복수로 표현했었음

Here, we present FloWaveNet, a flow-based approach that is an alternative to the real-time parallel generative model for raw audio synthesis. FloWaveNet requires only a single maximum likelihood loss, without any auxiliary loss terms, while maintaining stability in training. It features a simplified single-stage training scheme because it does not require a teacher network and can be trained end-to-end. The model is inherently parallel because of flow-based generation, which enables real-time waveform synthesis. FloWaveNet can act as a drop-in replacement for the WaveNet vocoder, which is used in a variety of text-to-speech architectures. Along with all the advantages described above, the quality and fidelity of samples from FloWaveNet are comparable to the two-stage models.

Currently, there is no official implementation of the aforementioned two-stage models available, and the performance of publicly accessible implementations does not match the result reported in their respective papers. In addition to our FloWaveNet, we present an open source implementation of the Gaussian IAF model that outperforms current public implementations, along with the first comparative study with the previous parallel synthesis model on a publicly available speech dataset. The code and samples for all models, including FloWaveNet, are publicly available\footnote{\url{https://github.com/ksw0306/FloWaveNet}}\footnote{\url{https://github.com/ksw0306/ClariNet}}\footnote{\url{https://ksw0306.github.io/flowavenet-demo}}.

Our major contributions are as follows:
\begin{itemize}
  \item We propose FloWaveNet, a new flow-based approach for parallel waveform speech synthesis which requires only a single maximum likelihood loss and an end-to-end single-stage training, in contrast to previous two-stage approaches.
  \item We show the difficulty of generating realistic audio with the two-stage approach without using the auxiliary loss terms, whereas the training of FloWaveNet is greatly simplified and stable throughout iterations.
  \item We present an open source implementation of FloWavenet and the Gaussian IAF that outperforms publicly available implementations, along with the first comparative study between methods using a publicly available speech dataset.
\end{itemize}
% 108에서 the difficulty를 강조하려고 앞으로 뺀 거같은데 표현 괜춘한지 with ~ without ~ 이런거
% 109에서 FWN에 the는 빼고 Gaussian IAF의 the는 안 뻈네?

The rest of this paper is organized as follows: In Section \ref{relatedworkwavenet}, we provide a summary of the original WaveNet and the speed bottleneck for real-world applications. Section \ref{relatedworkparallel} provides core backgrounds (IAF and \textit{probability density distillation}) of the recently proposed parallel speech synthesis method and we describe two previous works and their limitations. Section \ref{fwn} presents our FloWaveNet model and Section \ref{experiments} shows experimental details. We provide crowd-sourced mean opinion score (MOS) results and further analysis on the behaviors of each models in Section \ref{results}.

\section{Related Work} 

\subsection{WaveNet} \label{relatedworkwavenet}

WaveNet \cite{van2016wavenet} is a generative model that estimates the probability distribution of raw audio, and it can synthesize state-of-the-art fidelity speech and audio. The model decomposes the joint probability distribution of the audio signal $x_{1:T}$ into a product of conditional probabilities as follows:
\begin{equation}
\label{eq1}
P_{X}(x_{1:T}) = \prod_{t}P_{X}(x_t|x_{<t}).
\end{equation}
%% 초기 논문(Origianl WaveNet)에서는 각 시점마다 conditional probability를 모델링하기 위해 raw audio를 8-bit quantize하여 256-categorical distribution을 사용하였다. 이후 (Tacotron 2, Parallel WaveNet, ClariNet)에는 더 high fidelity audio를 생성하기 위해 conditional probability를 discretized mixture of Logistic distribution이나 single Gaussian distribution으로 16-bit audio를 직접적으로 모델링하였다.
The model estimates the joint probability by using causal dilated convolutions, which can successfully model the long-term dependency of the conditional probabilities. The original WaveNet architecture used an 8-bit quantized signal and estimated the conditional probabilities via a 256-way categorical distribution by using a softmax function. Subsequent studies \cite{oord2017parallel, ping2018clarinet} replaced the categorical distribution with a discretized mixture of logistics (MoL) or single Gaussian distribution directly by using 16-bit audio for a higher-fidelity sound.

%% WaveNet 구조는 fully convolutional network로 이루어져 있어 teacher forcing을 이용하여 모든 timestep에 대해서 parallel하게 학습이 가능하며, 아주 realistic한 audio sample을 합성할 수 있다. (RNN 들먹이기?) 하지만, WaveNet은 inference시에 autoregressive model의 특성인 ancestral sampling을 해야하므로 합성 속도가 느리다는 단점이 있다. (or 합성 속도가 느리다.)
The WaveNet structure is a fully convolutional network without recurrence, which can enable an efficient parallel probability estimation for all time steps and parallel training, via a teacher-forcing technique. However, because of the autoregressive nature of the model, WaveNet can only use the ancestral sampling scheme, which runs a full forward pass of the model for each time step during inference, which is a major speed bottleneck in the audio synthesis. The official implementation generates 172 samples per second, which is approximately 140 times slower than real-time for 24kHz audio \cite{oord2017parallel}.

\begin{figure*}[ht]
\begin{center}
\centerline{\includegraphics[width=\textwidth]{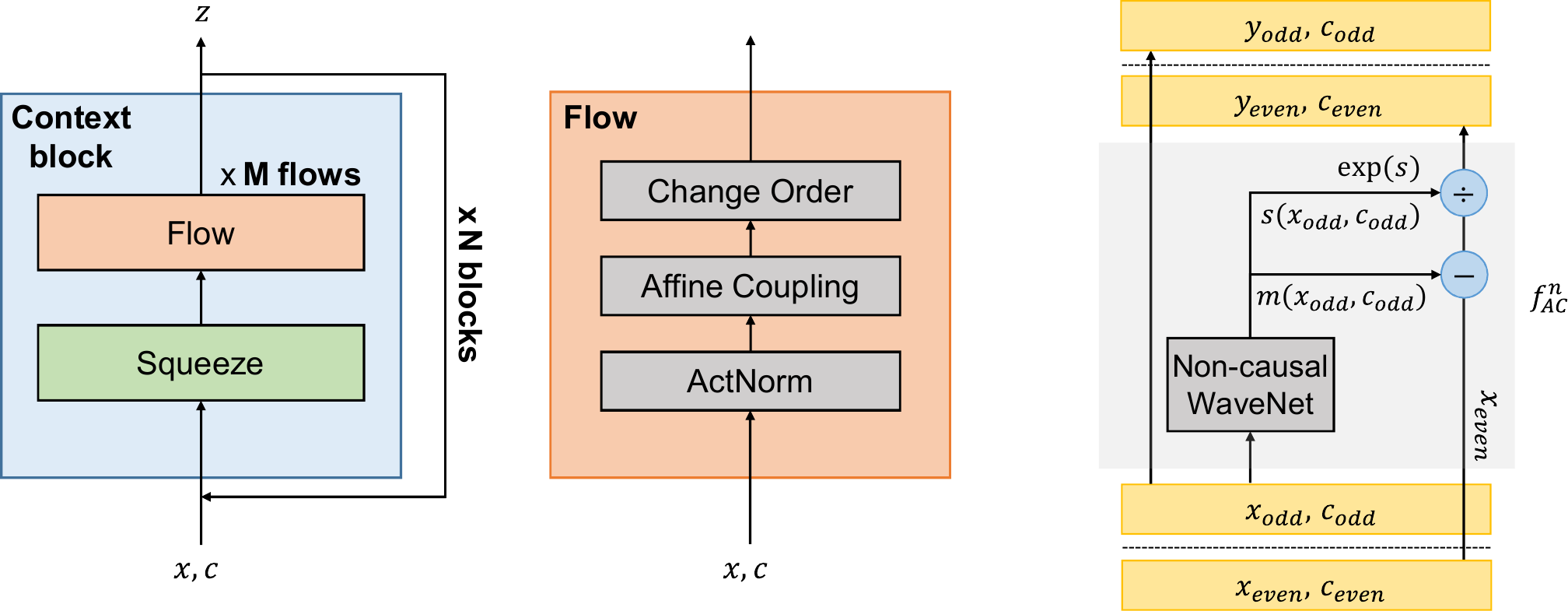}}
\caption{Schematic diagram of FloWaveNet. Left: an entire forward pass of the FloWaveNet consisting of N context blocks. Middle: an abstract diagram of the flow operation. Right: a detailed version of the affine coupling operation.}
\label{figure1}
\end{center}
\end{figure*}

\subsection{Parallel Speech Synthesis Models} \label{relatedworkparallel}

%% 이를 위해 이전의 Parallel Speech Synthesis Model들은 parallel sampling이 가능한 Inverse Autoregressive Flow(IAF)를 도입하였다. IAF는 식 (#)과 같이 simple distribution으로부터 뽑은 random noise를 audio data distribution으로 mapping 시키는 변환이다. IAF를 사용해서 주어진 audio data에 대한 likelihood를 estimate하는 것은 sequential computation이 필요하여 비효율적으로, maximum likelihood training 대신 pre-trained WaveNet을 이용하는 Probability density distillation 방식을 통해 IAF를 학습한다. (혹은 IAF를 maximum likelihood를 통해 학습하는 것은 sequential computation이 필요하여 비현실적(?or 비효율적)으로 대신 기존에 학습한 WaveNet을 이용한 Probability density distillation을 사용하여 학습한다.) 이는 given condition(e.g. mel spectrogram, linguistic features, ...)에 대해서 IAF로 변환한 분포 P_S와 pre-trained WaveNet가 estimate한 분포 P_T사이의 KL Divergence를 최소화하는 방식으로 IAF가 well-trained autoregressive WaveNet과 유사하게 density estimation을 할 수 있게 학습시킨다.
The main objective of parallel speech synthesis models \cite{oord2017parallel, ping2018clarinet} is to endow the system with the ability to synthesize audio in parallel at all time steps during inference, while also ensuring sound fidelity and quality comparable to the original autoregressive WaveNet. To this end, parallel speech synthesis models incorporate IAF \cite{kingma2016improved}, which is a special variant of normalizing flows \cite{rezende2015variational}. IAF transforms a random noise from a simple distribution into a complex data distribution via parallel sampling from the latent variables.

However, training IAF parameters via a maximum likelihood estimation requires sequential computations for each time step, which is computationally demanding and impractical. Instead, parallel speech synthesis models utilize a \textit{probability density distillation} training strategy \cite{oord2017parallel} to circumvent the issue, which enables the parallel IAF training. The strategy leverages a pre-trained WaveNet as a teacher and a student IAF is trained to minimize the KL divergence between the estimation from the student IAF and the teacher WaveNet, given conditions (e.g., mel spectrogram, linguistic features). The student IAF is optimized such that it can perform the density estimation by closely following the well-trained autoregressive WaveNet.

% Parallel WaveNet은 pre-trained WaveNet의 분포 P_T를 Mixture of Logistic distribution을 사용하였기 때문에 KL diergence에 대한 closed form이 없으므로 아래 식과 같이 KL divergence를 Monte Carlo estimation하여 minimize하였다. 따라서 모델을 안정적으로 학습시키기 위해서는 IAF로부터 나온 분포에서 더 많은 샘플을 뽑아서 estimated KL divergence를 구해야 한다. 
Parallel WaveNet \cite{oord2017parallel} uses MoL distribution for the pre-trained teacher WaveNet, which does not have a closed-form solution for the KL divergence. The model instead approximates the KL divergence via Monte Carlo estimation, which employs a trade-off between stability and speed, in which the model should generate enough samples from IAF to ensure stabilized training.
%% 수식 (넣을지 말지) 여기 수식 복잡

%% ClariNet 논문에서는 Parallel WaveNet에서의 approximation에 의한 불안정한 학습을 해결하기 위해 pre-trained WaveNet과 IAF의 output distribution을 모두 single Gaussian 분포로 두어 KL divergence의 closed form을 minimize하는 방식을 제안하였다. 추가적으로 high peak distribution인 P_T와 IAF로 변환한 분포 P_S의 standard deviation의 차이에 의해 생기는 numerical issue를 해결하기 위해 KL divergence 식을 식 (3)과 같이 변경하여 최적화하였다. 
ClariNet \cite{ping2018clarinet} suggested an alternative single Gaussian formulation for the KL divergence, which is closed-form and mitigates the instability of the Monte Carlo approximation of Parallel WaveNet. ClariNet additionally regularized the KL divergence to compensate for numerical stability when the difference in the standard deviation between a high peak teacher distribution $P_T = \mathcal{N}(\mu_T, \sigma_T^2)$ and the $P_S = \mathcal{N}(\mu_S, \sigma_S^2)$ from the student IAF becomes large, as follows:

\begin{equation}
\label{eq2}
KL(P_{S}||P_{T}) = \log{\frac{\sigma_{S}}{\sigma_{T}}} + \frac{\sigma_{S}^{2} - \sigma_{T}^2 + (\mu_{T} - \mu_{S})^{2}}{2\sigma_{T}^{2}},
\end{equation}
\begin{equation}
KL_{reg}(P_{S}||P_{T}) = KL(P_{S}||P_{T}) + \lambda |\log\sigma_{T} - \log\sigma_{S}|^{2}.
\end{equation}

%% 하지만, 위의 두 가지 parallel speech synthesis models들은 위의 probability density distillation loss만으로는 teacher의 특정 distribution에 collapsing되기 때문에, 추가적인 loss를 필요로 한다. Parallel WaveNet과 ClariNet 모두 오디오의 합성 퀄리티를 높이기 위하여 IAF로 sampling한 synthesized_audio와 original audio사이에 spectral disctance를 줄이는 추가적인 loss를 사용하였다. Parallel WaveNet에서는 이것을 제외하고도 perceptual loss, contrastive loss를 모두 조합하여 모델을 학습시켰다.
However, both of the aforementioned approaches for parallel speech synthesis require heavily-engineered auxiliary losses for the stabilized training because if the student IAF model uses only probability density distillation, it collapses to a certain mode of the teacher WaveNet distribution. Parallel WaveNet and ClariNet both used an additional spectral distance loss between the synthesized and original audio to ensure realistic sound quality. In addition, Parallel WaveNet further employed extra losses for additional improvements of the model, such as perceptual loss and contrastive loss.

\section{FloWaveNet} \label{fwn}

Here, we describe FloWaveNet, which learns to maximize the exact likelihood of the data while maintaining the ability of real-time parallel sampling, as an alternative to the two-stage training from the related work. FloWaveNet is a hierarchical architecture composed of context blocks as a highest abstract module and multiple reversible transformations inside the context block, as illustrated in Figure~\ref{figure1}.

\subsection{Flow-based generative model}
FloWaveNet is a flow-based generative model using normalizing flows \cite{rezende2015variational} to model raw audio data. Given a waveform audio signal $x$, assume that there is an invertible transformation $f(x): x \longrightarrow z $ that directly maps the signal into a known prior $P_Z$. We can explicitly calculate the log probability distribution of $x$ from the prior $P_Z$ by using a change of variables formula as follows:

\begin{equation}
\label{eq4}
\log{P_{X}(x)} = \log{P_{Z}(f(x)))} + \log{\det(\frac{\partial{f(x)}}{\partial{x}})}.
\end{equation}

The flow-based generative model can realize the efficient training and sampling by fulfilling the following properties: $(i)$ calculation of the Jacobian determinant of the transformation $f$ should be tractable in equation \eqref{eq4}, $(ii)$ mapping random noise $z$ into audio sample $x$ by applying the inverse transformation $x = f^{-1}(z)$ should be efficient enough to compute. Note that the parallel sampling becomes computationally tractable only if property $(ii)$ is satisfied.
% Hmm the 안 들어가는게 맞나...? 앞에서 (ii) 한 건디

To construct a parametric invertible transformation $f$ that fulfills both properties, FloWaveNet uses affine coupling layers suggested in real NVP \cite{dinh2016density}. To model the data using a transformation $f$ that is complex and flexible enough for audio, FloWaveNet stacks multiple flow operations inside each block, comprising the WaveNet affine coupling layers $f_{AC}$ and activation normalization $f_{AN}$ as in Figure~\ref{figure1}. The log determinant of the transformation $f$ in Equation \eqref{eq4} can be decomposed into the sum of per-flow terms as follows:
%% Equation 대문자?

\begin{equation}
\label{eq5}
\log{\det(\frac{\partial{f(x)}}{\partial{x}})} = \sum_{n=1}^{MN}\log{\det(\frac{\partial{(f_{AC}^{n}\cdot{}f_{AN}^{n})(x)}}{\partial{x}})},
\end{equation}

where N and M are the number of blocks and flows, respectively.

The change of variables formula in Equation \eqref{eq4} holds for a conditional distribution as well. FloWaveNet can estimate the conditional probability density by incorporating any arbitrary context as the conditional information. In this study, we use the mel spectrogram as a local condition $c$ for the network to model the conditional probability $p(x|c)$, similar to WaveNet used as a vocoder in a common neural Text-to-Speech pipeline \cite{shen2018natural}.

\subsection{Affine Coupling Layer}

A typical flow-based neural density estimator focuses solely on the density estimation as its main objective. The neural density estimator family has the advantage of using a much more flexible transformation, such as masked autoregressive flow \cite{papamakarios2017masked} and transformation autoregressive network \cite{oliva2018transformation}. However, it only satisfies property $(i)$ and not $(ii)$, making it unusable for our purpose.

In contrast, the affine coupling layer is a parametric layer suggested in real NVP \cite{dinh2016density} that satisfies both $(i)$ and $(ii)$, and can sample $x$ from $z$ efficiently in parallel. The layer enables the efficient bidirectional transformation of $f$ by making the transformation function bijective while maintaining computational tractability. Each layer is the parametric transformation $f_{AC}^{n}: x \longrightarrow y$, which keeps half of the channel dimension identical and applies the affine transformation only on the remaining half, as follows:

\begin{equation}
\label{eq6}
y_{odd} = x_{odd},
\end{equation}
\begin{equation}
\label{eq7}
y_{even} = \frac{x_{even} - m(x_{odd}, c_{odd})}{\exp({s(x_{odd}, c_{odd})})},
\end{equation}

where $m$ and $s$ are a shared non-causal WaveNet architecture, and $c$ is the local condition (e.g., mel spectrogram) fed by WaveNet to model the conditional probability $p(x|c)$. 

Similarly, for the inverse transformation $(f_{AC}^{n})^{-1}: y \longrightarrow x$, we have:

\begin{equation}
\label{eq8}
x_{odd} = y_{odd},
\end{equation}
\begin{equation}
\label{eq9}
x_{even} = y_{even} \odot \exp{{s(y_{odd}, c_{odd})}} + m(y_{odd}, c_{odd}).
\end{equation}

Note that the forward and inverse transformations use the same architecture $m$ and $s$, thus the model satisfies property $(ii)$, which endows the system with the ability of efficient sampling from $f^{-1}$. The Jacobian matrix is lower triangular, and the determinant is a product of the diagonal elements, which satisfies property $(i)$:

\begin{equation}
\log{\det(\frac{\partial{f_{AC}^{n}(x)}}{\partial{x}})} = -\sum_{even}s(x_{odd}, c_{odd}).
\end{equation}

A single affine coupling layer does not alter half of the feature, keeping it identical. To construct a more flexible transformation $f$, FloWaveNet stacks multiple flow operations for each context block. After the affine coupling of each flow, the change order operation swaps the order of $y_{odd}$ and $y_{even}$ before feeding them to the next flow so that all channels can affect each other during subsequent flow operations. 
% 음 by keeping it identical --> , keeping it identical 음 후자 문법 잘 모름

\subsection{Context Block} \label{cb}

\begin{figure}[t]
\begin{center}
\centerline{\includegraphics[width=\linewidth]{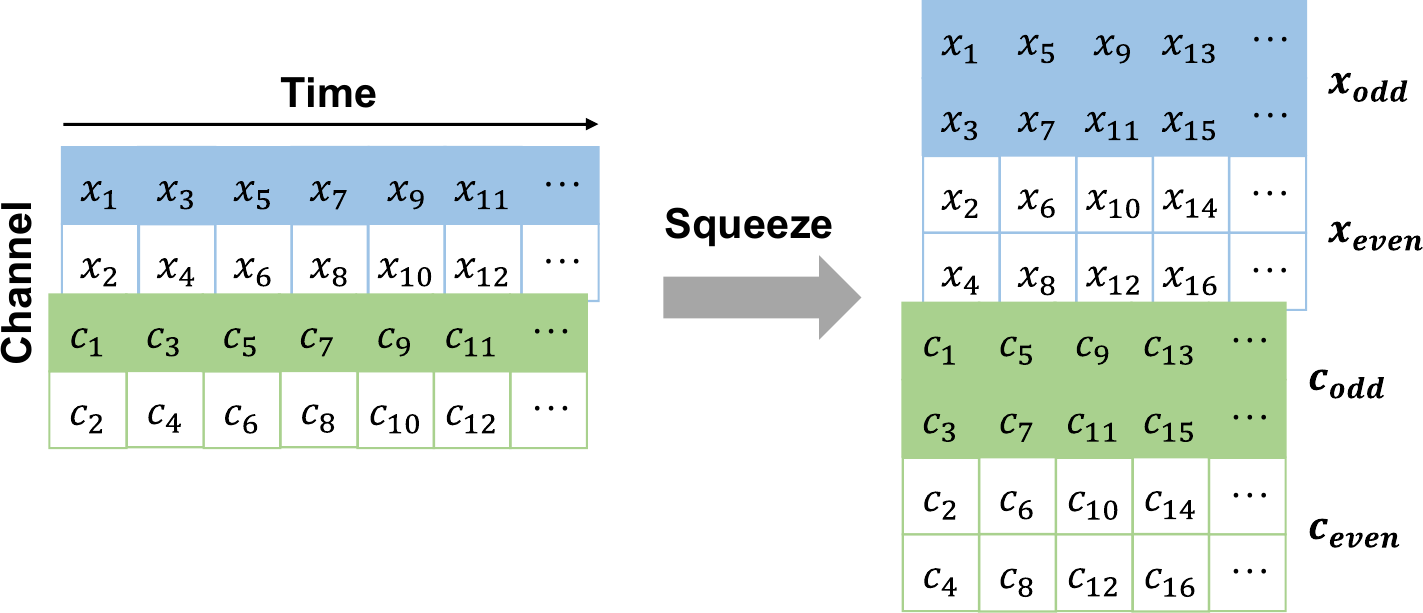}}
\caption{Squeeze operation used in the context block.}
\label{figure2}
\end{center}
\end{figure}

The context block is the highest abstraction module of FloWaveNet. Each context block consists of a squeeze operation followed by stacks of flow. The squeeze operation takes the data $x$ and condition $c$, then doubles the channel dimension $C$ by splitting the time dimension $T$ in half, as illustrated in Figure~\ref{figure2}. This operation doubles the effective receptive field per block for the WaveNet-based flow, which is conceptually similar to the dilated convolutions of the WaveNet itself. By applying the squeeze operation at the beginning of each context block, the upper-level blocks can have the potential to learn the long-term characteristics of audio, while the lower-level blocks can focus on high-frequency information. The flow operation inside the block contains an activation normalization, affine coupling layer, and change order operation, as described above.
% an을 맨 앞에만 붙이나

We employed a multi-scale architecture suggested in real NVP \cite{dinh2016density}. The multi-scale model factors out half of the feature channels, to be modeled as a Gaussian earlier after the pre-defined set of several context blocks and the remaining feature channels undergo subsequent context blocks. Thus, for the flows in the higher context blocks, the factored out channels act as identity mapping, whereas the remaining channels are further transformed. We used the same WaveNet-like architecture to estimate the mean and variance of the factored out Gaussian, using the remaining feature channels as inputs. This estimation strategy has minimal impact on the speed of synthesis compared to the pre-defined mean and variance of the factored out Gaussian, while producing higher-quality audio.

\subsection{Activation Normalization}

The activation normalization (ActNorm) layer suggested in Glow \cite{kingma2018glow} stabilizes the training of the network composed of multiple flow operations. The ActNorm layer $f_{AN}^{n}$ is a per-channel parametric affine transformation at the beginning of the flow. For $i$-th channel, we have:

\begin{equation}
f_{AN}^{n}(x_{i}) = x_{i} * s_{i} + b_{i},
\end{equation}
where $s$ and $b$ represent scale and bias for each channel. 

Note that $f_{AN}^{n}$ is a volume-changing operation, thus the log-determinant of the operation is computed as follows:

\begin{equation}
\log{\det(\frac{\partial{f_{AN}^{n}(x)}}{\partial{x}})} = T * \sum_{i=1}^{C}\log |s^{i}|.
\end{equation}

The layer performs data-dependent initialization of the trainable parameters $s$ and $b$ by scaling the activation channel-wise to have zero mean and unit variance for the first given batch of data.

\section{Experiments} \label{experiments}

We trained the model using the LJSpeech dataset \cite{ljspeech17}, which is a 24-hour waveform audio set of a single female speaker with 13,100 audio clips and a sample rate of 22kHz. We randomly extracted 16,000 sample chunks and normalized them to $[-1, 1]$ as the input. For local conditioning with the mel spectrogram construction, we used a preprocessing method from Tacotron 2 \cite{shen2018natural}. The generated 80-band mel spectrogram is used by the network to estimate the conditional probability.

We reproduced results from the original autoregressive WaveNet \cite{van2016wavenet} and the Gaussian IAF of ClariNet \cite{ping2018clarinet} as baselines. All models are trained under the mel spectrogram condition. We used an Adam optimizer \cite{kingma2014adam} with a learning rate of $10^{-3}$ for all models, identically to the ClariNet training configuration. We scheduled the learning rate decay by a factor of 0.5 for every 200K iterations. We used NVIDIA Tesla V100 GPUs with a batch size of 8 for all models.

\subsection{Autoregressive WaveNet} \label{aw}
We trained two autoregressive WaveNet models, one with the MoL and the other with a single Gaussian, as the output distribution. For the MoL WaveNet, we trained the best-performing autoregressive WaveNet from Tacotron 2 \cite{shen2018natural}, with the exact configuration from the paper, which is a 24-layer architecture with four 6-layer dilation cycles. For the single Gaussian WaveNet, we trained the 20-layer model with the same configuration used in ClariNet \cite{ping2018clarinet}. We trained the models for a total of 1M iterations. 

\subsection{Gaussian Inverse Autoregressive Flow (IAF)}
For the Gaussian IAF from ClariNet, we used the best-performing pre-trained single Gaussian autoregressive WaveNet from subsection \ref{aw} as the teacher network for the probability density distillation. We used the transposed convolution parameters from the teacher network without further tuning to upsample the mel spectrogram condition.

The student network has an architecture similar to the teacher network. It has a 60-layer architecture with six stacks of IAF modules, each of them with a 10-layer dilation cycle, which is the same configuration that is used in ClariNet \cite{ping2018clarinet}. We trained the model for a total of 500K iterations.

The ClariNet training requires a regularized KL divergence loss and an auxiliary spectrogram frame loss. In addition to the standard training, we performed an analysis of the impact of each loss by additionally training models with only one of the two losses.

\subsection{FloWaveNet}

FloWaveNet has 8 context blocks. Each block contains 6 flows, which results in a total of 48 stacks of flows. We used the affine coupling layer with a 2-layer non-causal WaveNet architecture \cite{van2016wavenet} and a kernel size of 3 for each flow. We used the multi-scale architecture described in \ref{cb}, in which we factored out half of the feature channels as a Gaussian after 4 context blocks, and we estimated the mean and variance of the Gaussian using the 2-layer WaveNet architecture, identically to the affine coupling layer.

We used 256 channels for a residual, skip, and gate channel with a gated tanh activation unit for all of the WaveNet architecture, along with the mel spectrogram condition. The weights for the last convolutional layer of the affine coupling are initialized with zero, so that it simulates identity mapping during the initial stage of training, which reportedly showed a stable training result. \cite{kingma2018glow}

We induced the model to learn the long-term dependency of audio by stacking many context blocks, instead of increasing the dilation cycle of the affine coupling. We trained the model with a single maximum likelihood loss without any auxiliary terms for 700K iterations.

\section{Results and Analysis} \label{results}
Our results with the sampled audio for all models are publicly available as described in Section \ref{intro}. 

The Gaussian IAF and FloWaveNet generate the waveform audio by applying the normalizing flows, using random noise samples as inputs. We set the standard deviation (\textit{i.e.}, temperature) of the prior below 1, which generated relatively higher-quality audio. We chose a temperature of 0.8 for FloWaveNet as the default, which empirically showed the best sound quality. 

\subsection{Model Comparisons}

\begin{table}[t]
\caption{Comparative mean opinion score (MOS) results with 95\% confidence intervals and conditional log-likelihoods (CLL) on test set.}
\label{mos-result}
\vskip 0.15in
\begin{center}
\begin{small}
\begin{sc}
\begin{tabular}{l|l|l}
\toprule
\textbf{Methods} & \textbf{5-scale MOS} & \textbf{Test CLL}\\
\midrule
Ground Truth    & 4.67$\pm$ 0.076 &\\
MoL WaveNet & 4.30$\pm$ 0.110 & 4.6546\\
Gaussian WaveNet   & 4.46$\pm$ 0.100 & 4.6526\\
Gaussian IAF    & 3.75$\pm$ 0.159 & \\
FloWaveNet     & 3.95$\pm$ 0.154 & 4.5457\\
\bottomrule
\end{tabular}
\end{sc}
\end{small}
\end{center}
\vskip -0.1in
\end{table}

\begin{table}[t]
\caption{Training iterations per second and inference speed comparison. The values for WaveNet and Parallel WaveNet are from \cite{oord2017parallel}, and the others are from our implementation.}
\label{speeds}
\vskip 0.15in
\begin{center}
\begin{small}
\begin{sc}
\begin{tabular}{l|l|l}
\toprule
\textbf{Methods} & \textbf{Iter/sec} & \textbf{Samples/sec}\\
\midrule
WaveNet & N/A & 172 \\  
Parallel WaveNet   & N/A & 500K\\
Gaussian WaveNet  & 1.329 & 44\\
Gaussian IAF    & 0.636 & 470K\\
FloWaveNet     & 0.714 & 420K\\
\bottomrule
\end{tabular}
\end{sc}
\end{small}
\end{center}
\vskip -0.1in
\end{table}
%% reference를 WaveNet이랑 Parallel WaveNet에 달아야 함

Table~\ref{mos-result} presents comparative results from a subjective 5-scale MOS experiment via Amazon Mechanical Turk, along with an objective conditional log-likelihoods (CLL) on test set. In general, the autoregressive WaveNet (MoL or Gaussian) received the highest CLL and MOS, which are closest to the ground truth sample. Between the autoregressive models, the Gaussian WaveNet performed slightly better than the MoL WaveNet, which is consistent with the result from ClariNet \cite{ping2018clarinet}.

The parallel speech synthesis models showed a slightly degraded audio quality compared to the autoregressive model, as also objectively evidenced by the lower test CLL. FloWaveNet showed a better performance evaluation result compared to that of our reproduced version of the Gaussian IAF. For the Gaussian IAF, there was an audible white noise throughout the samples which might negatively affect the evaluation of the audio quality compared to other models. FloWaveNet did not incur the white noise and had a clear sound quality unlike the Gaussian IAF. But instead, there was a periodic artifact perceived as a trembling voice, which varied for different temperatures and we discuss it in Section \ref{Temperature}.

Overall, the sound quality of FloWaveNet was comparable to the previous approaches as well as having the advantages of the single maximum likelihood loss and the single-stage training. Note that although the autoregressive WaveNet showed the highest fidelity, it requires slow ancestral sampling. Table \ref{speeds} shows the inference speed comparison for each model. FloWaveNet can generate the 22,050Hz audio signal approximately 20 times faster than real-time, which is similar in magnitude to the reported speed results from the Parallel WaveNet \cite{oord2017parallel}.

Our reimplemented Gaussian IAF from the ClariNet model achieved a sampling speed approximately similar to FloWaveNet. Note that the difference in sampling speed between various parallel speech synthesis architectures (including our FloWaveNet model) is largely from a selection of the number of channels of convolutional layers. The Parallel WaveNet and ClariNet used 64 channels for the reported model architecture in their respective papers. The experimental results from FloWaveNet used 256 channels as our default settings. However, one can also construct a smaller version of FloWaveNet with 128 channels which results in a slightly degraded performance but with twice the sampling speed.

\begin{figure*}[ht]
\begin{center}
\centerline{\includegraphics[width=\textwidth]{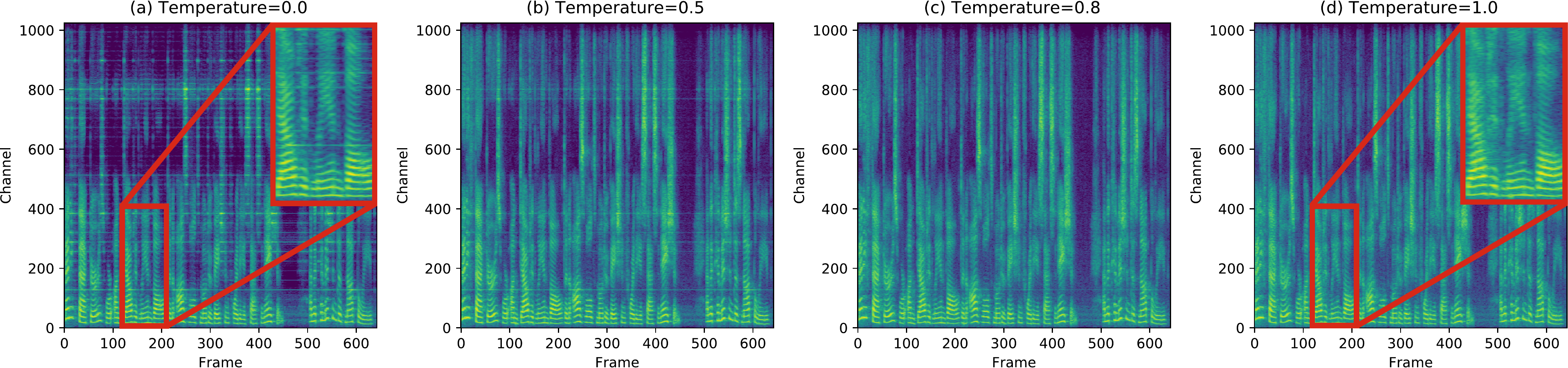}}
\caption{Spectrogram visualization of FloWaveNet samples with different temperatures.}
\label{figure3}
\end{center}
\end{figure*}

We also reported the number of training iterations per second for Gaussian WaveNet, Gaussian IAF, and FloWaveNet with a single NVIDIA Tesla V100 GPU in Table \ref{speeds}. Note that we chose this metric as the most relevant and dataset-independent speed benchmark for training. As the Gaussian IAF requires two-stage training procedure, its total training wall-clock time is the sum of training Gaussian WaveNet and Gaussian IAF: Each of them takes 8.7 and 9.1 GPU days for convergence on the LJSpeech with the aforementioned training iterations, respectively. Training FloWaveNet takes 11.3 GPU days, demonstrating that FloWaveNet has a practical merit of faster convergence since it does not necessitate the well-trained teacher network.

\subsection{Temperature Effect on Audio Quality Trade-off} \label{Temperature}

The empirical experiments presented in Glow \cite{kingma2018glow} included adjusting the temperature of the pre-defined known prior. Glow maximized the perceptive quality of generated images by drawing random samples with the temperature below 1 and then applying inverse transformation to the random samples. We performed a similar empirical study on the effects of the temperature in terms of audio.

The flow-based generative models can perform an interpolation between two data points in the latent space via the latent variables. We interpret the temperature effect from the perspective of a latent traversal, where lowering the temperature $T$ corresponds to performing the latent traversal between a random sample $z$ from the known prior and a zero vector $O$:

\begin{equation}
P_{\hat{Z}}(\hat{z}) = \mathcal{N}(0, T^{2}) \Longleftrightarrow \hat{z} = (1 - T) * O + T * z.
\end{equation}

Figure \ref{figure3} represents the visualized spectrogram of the latent traversal with temperatures of 0, 0.5, 0.8, and 1.0. The spectrogram in Figure \ref{figure3} (a) corresponds to the audio using the zero vector. We can see that it exhibits many horizontal lines, which correspond to constant noises for multiple pitches, while instead, the zero vector tends to model the harmonic structure of the speech more strongly with high resolution, as can be seen in the zoomed-in view. In contrast, the spectrogram in Figure \ref{figure3} (d) shows that the temperature of 1.0 generates a high-quality audio without the constant noise artifacts. However, it is harder to capture the harmonic structure of the speech content using the high temperature, which is translated into the trembling voice.
% 398 figure 같이 찍어서 안 보냈더니 혼쭐남 we can see that is exhibits 여기서 it이 뭐를 가르키는지 모를 듯

From the latent traversal analysis, we can see that there exists a trade-off between high-fidelity quality and the harmonic structure of the speech. We could generate high-quality speech via lowering the temperature to minimize the trembling, while also ensuring that the constant noise artifacts are inaudible and not discomforting. We maximized the perceptive audio quality through the empirical post-processing approach of temperature optimization in this work, and further research on improving flow-based generative architectures would be a promising direction for future work.

\begin{figure*}[ht]
\begin{center}
\centerline{\includegraphics[width=\textwidth]{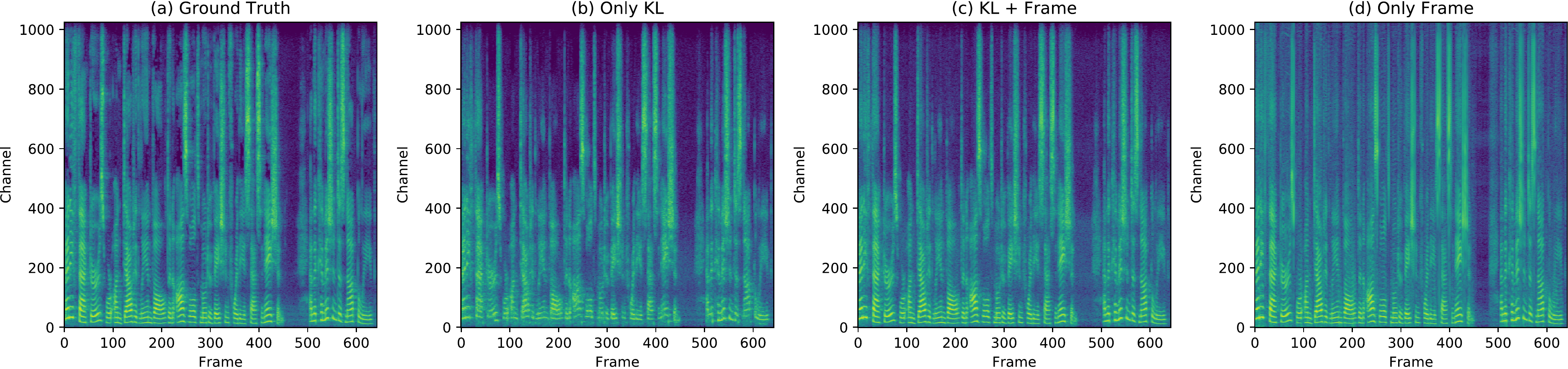}}
\caption{Spectrogram visualization of ClariNet samples with different loss configurations.}
\label{figure4}
\end{center}
\end{figure*}

\begin{table}[t]
\caption{KL divergence results on test data, estimated between each Gaussian IAF and the teacher WaveNet.}
\label{nll}
\vskip 0.15in
\begin{center}
\begin{small}
\begin{sc}
\begin{tabular}{lc}
\toprule
Method: Gaussian IAF & KL divergence\\
\midrule
Only KL & 0.040 \\
KL + Frame   & 0.134 \\
Only Frame    & 1.378 \\
\bottomrule
\end{tabular}
\end{sc}
\end{small}
\end{center}
\vskip -0.1in
\end{table}

\subsection{Analysis of ClariNet Loss Terms} \label{clarinetlossanalysis}

Parallel WaveNet \cite{oord2017parallel} and ClariNet \cite{ping2018clarinet} are parallel waveform synthesis models based on IAF, as discussed earlier. They use KL divergence in tandem with additional auxiliary loss terms to maximize the quality of the sampled audio. Here, we present an empirical analysis of the role of each term. We decomposed the two losses of ClariNet and trained separate Gaussian IAF models with only one of the losses: the KL divergence or the spectrogram frame loss.

Table \ref{nll} shows a quantitative analysis on the KL divergence between the estimation from each Gaussian IAF training method and the pre-trained teacher WaveNet, given the test data mel spectrogram condition. Figure \ref{figure4} represents the spectrogram examples generated by each method. The waveform audio samples corresponding to each method are also publicly available at the provided webpage. 
% 433에 as condition에서 as를 빼버렸는데 상관없으려나

We can clearly see from Table \ref{nll} that for the Gaussian IAF, using only the KL divergence loss showed the best reported metric on the test data, which is a direct result from the explicit optimization. However, samples generated by the KL-only training, although phonetically perceptible, are low-volume and sound distorted. Figure \ref{figure4} (b) shows that the signal has low-energy across all samples compared to the ground truth and that the model has limitations on estimating the harmonics of the speech signal in the mid-frequency range ($e.g.$, 2$\sim$5kHz). Not only does this show that a good KL divergence result does not necessarily mean a realistic-sounding audio, but also the Gaussian IAF is prone to mode collapse problem when using the probability density distillation loss only.

The model trained with only the spectrogram frame loss showed relatively fast estimation of the acoustic content of the original audio earlier in training ($e.g.$, ~10K iterations). The spectrogram in Figure \ref{figure4} (b) more closely resembles the ground truth spectrogram because the frame loss directly targets the resemblance in the frequency domain. However, the generated samples showed a high amount of noise and the noise did not diminish without the KL divergence loss in the remaining training iterations.

\begin{table}[t]
\caption{Mean opinion score (MOS) results with non-causal and causal variants of FloWaveNet.}
\label{causalitymos}
\vskip 0.15in
\begin{center}
\begin{small}
\begin{sc}
\begin{tabular}{lc}
\toprule
Method: FloWaveNet & 5-scale MOS\\
\midrule
Non-causal & 3.95 $\pm$ 0.154 \\
Causal & 3.36 $\pm$ 0.134 \\
\bottomrule
\end{tabular}
\end{sc}
\end{small}
\end{center}
\vskip -0.1in
\end{table}

Ideally, the probability density distillation enables the student network perfectly mimic the distribution estimated by the teacher WaveNet. However, the distilled model trained only by the KL divergence shows its limitations of covering every mode of the teacher model, as evidenced by this study and by previous works \cite{oord2017parallel}. This shortcoming can be alleviated by incorporating the spectrogram frame loss into the distillation process, as shown by the Figure \ref{figure4} (c), where the model starts to appropriately track the harmonics and estimate the original amplitude. Thus, the Gaussian IAF training requires both complementary loss terms for realistic-sounding speech synthesis.

\subsection{Causality of WaveNet Dilated Convolutions}

The original WaveNet achieved autoregressive sequence modeling by introducing causal dilated convolutions. However, for FloWaveNet, the causality is no longer a requirement because the flow-based transformation is inherently parallel in either direction. We performed an ablative study by comparing both approaches, and the non-causal version of FloWaveNet exhibited better sound quality, as can be seen in Table~\ref{causalitymos}. This is because the non-causal version of FloWaveNet has the benefit of observing the mel spectrogram condition both forward and backward in its receptive field.

\section{Conclusion}
In this paper we proposed FloWaveNet, a flow-based generative model that can achieve a real-time parallel audio synthesis that is comparable in fidelity to two-stage approaches. Thanks to the simplified single loss function and single-stage training, the model can mitigate the need for highly-tuned auxiliary loss terms while maintaining stability of training which is useful in practical applications. Our results show that the flow-based generative model is a promising approach in speech synthesis domain which shed light on new research directions.

\section*{Acknowledgements}
This work was supported by the National Research Foundation of Korea (NRF) grant funded by the Korea government (Ministry of Science and ICT) [No. 2018R1A2B3001628], Hyundai Motor Company, AIR Lab (AI Research Lab) in Hyundai Motor Company through HMC-SNU AI Consortium Fund, Samsung Electronics Company, and Brain Korea 21 Plus Project in 2019.

% \textbf{Do not} include acknowledgements in the initial version of
% the paper submitted for blind review.

% If a paper is accepted, the final camera-ready version can (and
% probably should) include acknowledgements. In this case, please
% place such acknowledgements in an unnumbered section at the
% end of the paper. Typically, this will include thanks to reviewers
% who gave useful comments, to colleagues who contributed to the ideas,
% and to funding agencies and corporate sponsors that provided financial

% % In the unusual situation where you want a paper to appear in the
% % references without citing it in the main text, use \nocite

\balance
\bibliography{main}
\bibliographystyle{icml2019}

\end{document}